\documentclass[conference]{IEEEtran}
\IEEEoverridecommandlockouts

\usepackage{cite}
\usepackage{amsmath,amssymb,amsfonts}
\usepackage{algorithmic}
\usepackage{graphicx}
\usepackage{textcomp}
\usepackage{url}
\usepackage{tabularray}
\usepackage{xcolor}
\usepackage{comment}
\usepackage[draft]{hyperref}
\begin{document}

\title{Are We Shooting Flies with Cannons? Trade-off Analysis for AI-based 5G Intrusion Detection\\

\thanks{This work was carried out within the NEST project $AIR^2$, which is partially supported by the Wallenberg AI, Autonomous Systems and Software Program (WASP) funded by the Knut and Alice Wallenberg Foundation.}
}

\author{
\IEEEauthorblockN{
Federica Uccello\IEEEauthorrefmark{1},
Simin Nadjm-Tehrani\IEEEauthorrefmark{1}
}
\IEEEauthorblockA{\IEEEauthorrefmark{1}
\textit{Department of Computer and Information Science} \\
\textit{Linköping University}\\
Sweden \\
\{federica.uccello, simin.nadjm-tehrani\}@liu.se
}
}
\maketitle

\begin{abstract}
The increasing adoption of Artificial Intelligence (AI) in network intrusion detection raises the question of whether complex and computationally expensive models are justified for this task. In this work, we investigate the trade-off between detection performance and computational cost for intrusion detection in 5G network telemetry.
We compare traditional machine learning (ML) models, including XGBoost as a representative of tree ensemble, and TabNet for tabular deep neural network (DNN), with a large language model (LLM) used as a general-purpose intrusion detector. The LLM is evaluated under both zero-shot and few-shot prompting configurations.
We evaluate the models in terms of detection performance, inference time, and CPU time as a proxy for energy efficiency. Using a relatively large available 5G dataset, we show that traditional ML models consistently achieve near-perfect detection performance with negligible inference time, while LLM-based approaches perform significantly worse and incur orders-of-magnitude higher CPU usage. Few-shot prompting improves recall, but at the cost of lower accuracy and further increased CPU time, without closing the performance gap.
These findings indicate that, for tabular intrusion detection in 5G networks, XGBoost offers a substantially better performance–cost trade-off than DNNs and LLMs, highlighting the importance of selecting models based on task suitability rather than increasing complexity.
\end{abstract}


\section{Introduction}
The latest developments in 5G and beyond have come at the cost of increased attack surface, motivating the development of new security solutions. According to recent ENISA threat landscape reports, the telecommunications sector remains one of the primary targets of cyber attacks \cite{enisa2025, enisa2024}.  In this context, Machine learning (ML) techniques are increasingly posed for adoption, particularly for analyzing network telemetry data represented in tabular form and handling the high volume and complexity of modern network traffic \cite{morocho2019machine}. 

Recently, the rapid advancement and widespread adoption of large language models (LLMs) have led to a growing interest in applying these models to a broad range of tasks in various domains, including cybersecurity \cite{ferrag2025generative,uccello2026mission}. 
Despite their promising capabilities, training and running such models require significant resources, raising concerns about their environmental impact, including increased electricity consumption, carbon emissions, and hardware resource utilization \cite{bolon2024review}.

In this work, we hypothesize that LLM-based approaches do not provide a favorable trade-off between detection performance and computational cost compared to traditional ML models, including tree ensembles and deep neural networks (DNNs), when applied to tabular 5G network intrusion detection.

The contributions of the paper are as follows:
\begin{itemize}
    \item We perform a comparative evaluation of representative tree ensembles, DNNs, and general-purpose LLMs for intrusion detection in 5G network telemetry.
    
    \item We analyze the impact of prompting strategies by comparing zero-shot and few-shot configurations for the LLM.
    
    \item We investigate trade-offs in terms of detection performance, inference time, and CPU time as a proxy for energy efficiency.

\end{itemize}

The remainder of this paper is organized as follows. Section \ref{sec:rw} overviews the related work; Section \ref{sec:bg} provides background concepts; Section \ref{sec:method} describes the aim and method;  Section \ref{sec:setup} presents the experimental setup; Section \ref{sec:result} reports the results, which are discussed in Section \ref{sec:disc}, along with threats to validity. Finally, Section \ref{sec:conclusion} concludes the paper with remarks on future directions.

\section{Related Work}
\label{sec:rw}
ML-based approaches have been widely adopted for network intrusion detection. 

Liu et al.\cite{liu2021fast} demonstrate that tree-based ensemble methods achieve high detection accuracy while maintaining low training and inference complexity across multiple benchmark datasets. Similarly, Waghmode et al.\cite{waghmode2025intrusion} provide an experimental comparison of several ML classifiers on standard intrusion detection datasets, showing that tree-based models can achieve high detection performance while exhibiting faster runtime than several neural network-based approaches. Zoppi et al.~\cite{zoppi2024anomaly} conducted an extensive comparative study, showing that tree ensemble models consistently outperform DNNs for anomaly-based intrusion detection on tabular data, even at a large scale. Our study extends this line of work by including LLM-based approaches and explicitly accounting for inference and computational cost, which is a main factor in energy consumption.

Our work is motivated by the fact that LLMs are increasingly being explored for cybersecurity applications, including network intrusion detection. Adjewa et al.~\cite{adjewa2025llmids} propose a transformer-based intrusion detection framework that captures contextual patterns in network traffic and adapts to previously unseen attacks through continuous learning mechanisms. Rezaei et al.~\cite{rezaei2025fedllmguard} integrate LLMs within a federated learning framework for anomaly detection in 5G networks, highlighting their potential for context-aware and privacy-preserving detection. 

However, the practical deployment of LLM-based approaches remains challenging due to their substantial computational and energy requirements. Argerich et al.~\cite{argerich2024measuring} show that LLM inference efficiency is strongly influenced by model size, architecture, batch size, and quantization, significantly impacting both inference time and energy consumption. Our work examines whether such computationally demanding models are justified for tabular 5G intrusion detection when compared with classical ML paradigms.

In our previous work \cite{uccello2026mission}, we showed that XGBoost-based models trained exclusively on benign data can achieve promising performance in detecting anomalies in 5G environments. There, we also used LLM to generate mitigation actions, validated against an expert-curated database with encouraging results. In contrast, the present work tests three different models with labeled data. 

To the best of our knowledge, existing work has not jointly evaluated detection performance and computational consumption when comparing traditional ML, DNN, and LLM-based approaches for tabular 5G intrusion detection. 

\section{Background}
\label{sec:bg}

This section provides a high-level overview of the model families considered in this work. For detailed theory, we refer the reader to the dedicated surveys\cite{mienye2022ensemble,borisov2022deep,kumar2024large}.

\subsection{Tree Ensemble Methods}
Tree ensemble methods are widely used for classification tasks involving tabular data. These approaches construct multiple decision trees and combine their predictions to improve generalization performance. By aggregating several base learners, ensemble methods typically achieve higher accuracy and better robustness than individual models, while also mitigating overfitting~\cite{mienye2022ensemble}.

\subsection{Deep Neural Networks}
DNNs are composed of multiple layers that learn hierarchical representations of input data. They have achieved strong performance across domains such as computer vision and natural language processing. However, their application to tabular data remains challenging. While there is no extensive understanding of why this is the case, it has been observed that the performance of DNNs on tabular tasks often depends heavily on feature preprocessing, architecture design, and training strategies  \cite{borisov2022deep}.

\subsection{Large Language Models}
LLMs are transformer-based architectures trained on large-scale data to learn general-purpose representations of language. Beyond natural language processing, LLMs have recently been adapted to other domains by encoding structured or semi-structured data into textual formats. However, their deployment in domain-specific applications remains challenging due to high computational requirements, limited interpretability, and the need for task-specific adaptation \cite{kumar2024large}.

\section{Method}
\label{sec:method}



This study attempts to add clarity in the multi-dimensional space relating to performance-inference time and performance-energy efficiency tradeoffs. In particular, whether LLMs have an edge over traditional ML models in the network telemetry domain. In order to study LLMs in this context, we also compare two different prompting strategies for LLMs.

As a representative of tree ensembles, we employ XGBoost \cite{chen2016xgboost}, a highly scalable gradient-boosted decision tree algorithm that has demonstrated strong performance in network intrusion detection tasks \cite{edozie2025artificial}. 

To provide a competitive deep learning baseline, we include TabNet \cite{zoppi2024anomaly}, a neural network architecture specifically designed for tabular data, which is the predominant format in ML-based intrusion detection.

For the LLM-based approach, we use Open-Mistral-7B\footnote{\url{https://docs.mistral.ai/}}. This model is selected due to its favorable performance-to-efficiency ratio among open-weight LLMs \cite{ferrag2025generative}. The LLM is deployed locally using \texttt{llama.cpp}\footnote{\url{https://github.com/ggml-org/llama.cpp}}, enabling CPU-based inference on the same hardware as the ML models. 
While LLMs are typically deployed with GPU acceleration or via cloud services in production, this setup allows us to evaluate all approaches under comparable resource conditions.

\section{Experimental Setup}
\label{sec:setup}

All experiments are implemented in Python and executed in a Debian-based virtual machine equipped with an Intel Core Ultra 7 155U CPU and 32 GB RAM\footnote{Libraries: scikit-learn, xgboost, pytorch-tabnet, llama.cpp}. Source code and data are available on GitLab\footnote{\url{https://gitlab.liu.se/feduc51/AI_Tradeoff}}. No other operations were executed on the machine during the experiments. 

\subsection{Dataset and Preprocessing}
\label{sub:data}

We use the 5G Network Intrusion Detection (5G-NIDD) dataset \cite{5gnidd}, which provides tabular network telemetry labeled as benign (39.3\%) or malicious (60.7\%). The dataset contains 1.2 million samples and 92 features extracted from traffic generated in a 5G experimental environment, including flows associated with Denial of Service (DoS) and port scanning attacks. We formulate the task as binary classification, mapping benign samples to class 0 and attacks to class 1. 

The data is split into training, validation, and test sets (72\%/8\%/20\%) using stratified sampling to preserve class distribution. Preprocessing is fitted on the training set only to avoid data leakage and includes removal of constant features, selection of numerical attributes, and feature selection using a filter-based method (\texttt{SelectKBest} with the ANOVA F-test, \texttt{f\_classif}). This method ranks features by their statistical correlation with the class label. As a result, the feature space is reduced to 20, including protocol information, packet statistics, routing characteristics, and TCP flag behavior. This dimensionality reduction ensures that LLM prompts remain concise, limiting input length, which is a good practice in prompt engineering \cite{levy2024same}. The same preprocessed data was used for all three models. 

Due to the high computational demand of LLM inference, the LLM-based evaluation is conducted on a stratified subset of 300 test samples, preserving the original class distribution to ensure representativeness. The same subset is used for both zero-shot and few-shot settings. 
 
\subsection{Intrusion Detectors}
Here, we briefly summarize the configuration of each model. The details can be found in the code. 

Both XGBoost and TabNet are trained on the same preprocessed training data, with the validation set used for hyperparameter tuning, performance monitoring, and overfitting mitigation. Final results are reported on the held-out test set only.
For XGBoost, class imbalance is handled through the \texttt{scale\_pos\_weight} parameter\footnote{See XGBoost user guide for definitions \url{https://xgboost.readthedocs.io/en/release_3.2.0/}}, computed from the training data.
For TabNet, early stopping is applied based on validation performance. 

To apply Mistral to tabular data, each sample is converted into a textual representation consisting of feature-value pairs. In all cases, the model is instructed through a system prompt to act as a binary classifier for 5G network telemetry and to reply with exactly one label, either \texttt{Benign} or \texttt{Malicious}, without additional explanation.

We evaluate two prompting strategies \cite{brown2020language}:
\begin{itemize}
    \item \textbf{Zero-shot (ZS):} each prompt contains a single sample.
    \item \textbf{Few-shot (FS):} each prompt includes two labeled examples, one benign and one malicious, sampled from the validation set, followed by the test sample to classify.
\end{itemize}


\subsection{Evaluation Strategy}
We evaluate the models along three dimensions:
\begin{itemize}

\item \textbf{Detection performance}: 
Standard classification metrics\footnote{See scikit-learn user guide for metrics definitions: \url{https://scikit-learn.org/stable/modules/model_evaluation.html}} are used to quantify the ability of each model to distinguish between benign and malicious samples correctly. These include accuracy, precision, recall, F1-score, and balanced accuracy (average recall across classes, selected to mitigate the effect of class imbalance). In addition, confusion matrix components (true positives, false positives, true negatives, and false negatives) are reported to provide a detailed breakdown of classification outcomes.

\item \textbf{Inference Time}: 
Prior work has shown that performance metrics alone provide an incomplete view of intrusion detection systems, as they do not capture how long an attack remains undetected \cite{puccetti2026detection}. Inference time is a major component when determining whether the time to detection for a given (ongoing) attack is decisive. 
Therefore, we incorporate inference time measurements in our experiments. 

We report the \textit{average inference time} to evaluate the speed with which the models raise a single alert. The average is calculated on the cardinality of the sample set used for testing. For the ML models, the entire test set was used, while the LLM was evaluated on a smaller subset. Since the subset of samples to generate prompts is selected to be representative of the entire test set (Section \ref{sub:data}), we believe the average per-sample provides a fair comparison. 

For completeness, we also report the total \textit{wall-clock time} for each model. For the ML models, this includes both training and the total inference time for all the samples. Since the LLMs are pre-trained, we are not able to provide a measure of the training time. Our experiments also exclude the time necessary for prompt engineering and deployment of the local LLM.

\item \textbf{CPU Time}:  
We use CPU time as a proxy for energy efficiency. CPU time is defined as the total processor time consumed by the process, expressed in seconds\footnote{See psutil user guide for metric definition: \url{https://psutil.readthedocs.io/stable/}}. Given that the average power draw for a CPU under a given load is known, energy usage can be estimated as \(E = P \times t\), where \(P\) is the power draw (in watts) and \(t\) is the execution time (in hours). 

In this context, CPU time provides an approximation of the energy consumed by the models together with the power consumption of the underlying system. Importantly, this metric is accumulated across multiple CPU cores utilized during execution (12 cores in the machine used for the experiments). As a result, CPU time reflects the total computational work performed by the system, rather than the elapsed wall-clock time defined above. 

For traditional ML models (XGBoost and TabNet), we measure the CPU time during both training (\textit{train CPU time}) and inference (\textit{test CPU time}). 

For the LLM, we report CPU time as the processor time consumed during prompt processing and response generation. This includes both the LLM inference process (hosted via \texttt{llama.cpp}), referred to in the following as \textit{test CPU time}, and the client-side execution responsible for issuing requests and handling responses. 

All measurements were conducted locally on the same hardware, and the LLM was not serving any external requests during evaluation, ensuring that the reported CPU usage reflects only the workload induced by the experiment.
\end{itemize}

\section{Result}
\label{sec:result}
\subsection{Detection Performance}
Table \ref{tab:performance} shows that both XGBoost and TabNet achieve near-perfect performance across all metrics. In contrast, the LLM-based approach performs significantly worse. The zero-shot configuration achieves a balanced accuracy of 0.67, while the few-shot configuration further degrades to 0.50, indicating a strong bias toward predicting the malicious class. Although few-shot prompting improves recall (reaching 1.00), this comes at the expense of an even stronger bias and a high false positive rate, with the model systematically flagging every sample as malicious.

\begin{table*}
\centering
\caption{Detection performance on the test set.}
\label{tab:performance}
\begin{tblr}{
  row{1} = {font=\bfseries},
  column{even} = {c},
  column{3} = {c},
  column{5} = {c},
  column{7} = {c},
  column{9} = {c},
  hline{1-2,6} = {-}{},
}
Model        & TN    & FP  & FN  & TP     & Acc  & Bal Acc & Prec & Rec  & F1   \\
XGBoost          & 95532 & 11  & 54  & 147577 & 0.99 & 0.99    & 0.99 & 0.99 & 0.99 \\
TabNet       & 95535 & 8   & 126 & 147505 & 0.99 & 0.99    & 0.99 & 0.99 & 0.99 \\
Mistral (ZS) & 89    & 29  & 75  & 107    & 0.65 & 0.67    & 0.79 & 0.59 & 0.67 \\
Mistral (FS) & 0     & 118 & 0   & 182    & 0.60 & 0.50    & 0.61 & 1.00 & 0.75 
\end{tblr}
\end{table*}

\subsection{Inference Time}
Table \ref{tab:inf} shows the average inference time per sample, and the total wall-clock time (including the training time for XGboost and TabNet). As reported, there are substantial differences in inference time across the models. Traditional ML models exhibit low inference time, with per-sample times on the order of microseconds.

In contrast, the LLM-based approach incurs several orders of magnitude higher inference time, requiring approximately 10 seconds per sample, and that excludes the time to formulate the prompts and select the examples provided. Additionally, the training time is not considered in the wall-clock time for the LLM. Considerations on typical LLM training times are in Section \ref{sec:disc}.

\begin{table}
\centering
\caption{Inference time comparison across models.}
\label{tab:inf}
\begin{tblr}{
  row{1} = {font=\bfseries},
  column{2} = {c},
  column{3} = {c},
  hline{1-2,6} = {-}{},
}
Model        & Avg Inference Time (s) & Wall-clock Time (s)                      \\
XGBoost      & $5.30 \times 10^-7$        & 75.88                                \\
TabNet       & $1.59 \times 10^-5$        & 831.71                                \\
Mistral (ZS) & 9.99                      & 2998.65
     \\
Mistral (FS) & 10.56                     & 3169.20
\end{tblr}
\end{table}

\subsection{CPU Time}

Table \ref{tab:cpu} shows the total CPU time for each model. Note that for the ML models, most CPU time is spent on training. To give an idea of how each operation contributes to the CPU time, we also report the train (where applicable) and test CPU time separately. The values are reported in seconds.

\begin{table}
\centering
\caption{CPU time comparison across models.}
\label{tab:cpu}
\begin{tblr}{
  row{1} = {font=\bfseries},
  column{even} = {c},
  column{3} = {c},
  hline{1-2,6} = {-}{},
}
Model        & Total CPU (s) & Train CPU (s) & Test CPU (s) \\
XGBoost      & 943.14        & 941.46           & 1.68            \\
TabNet       & 5319.27       & 5292.58          & 26.69            \\
Mistral (ZS) & 41720.50      & N/A              & 20859.85       \\
Mistral (FS) & 44168.18      & N/A              & 22083.72       
\end{tblr}
\end{table}

While XGBoost and TabNet require modest computational resources, the LLM-based approach consumes significantly higher CPU time, both for server and client. Notably, few-shot prompting further increases the CPU time without yielding corresponding performance improvements.

\begin{figure}
    \centering
    \includegraphics[width=0.8\linewidth]{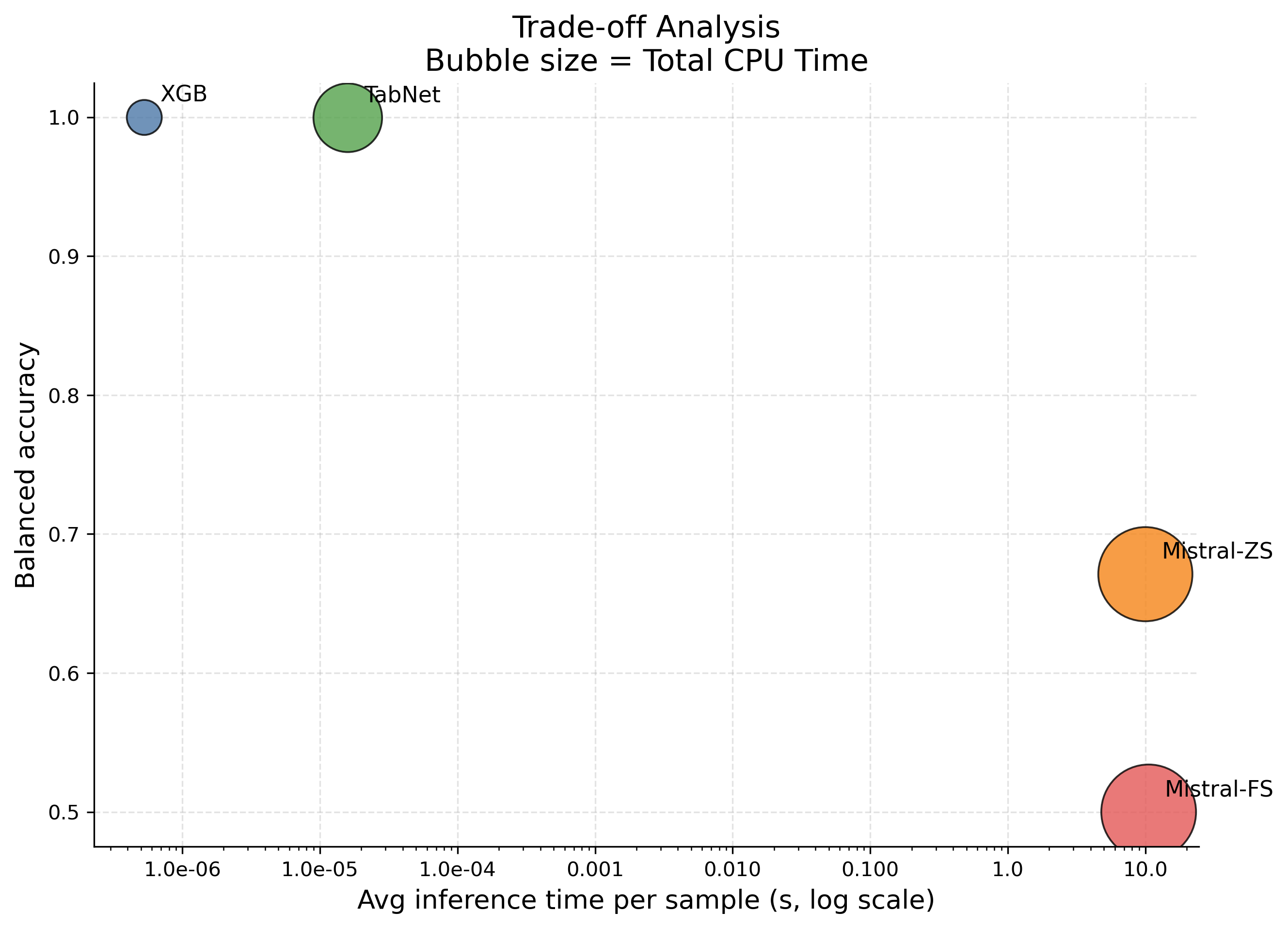}
    \caption{Trade-off between detection performance, average inference time, and CPU time across models.}
    \label{fig:tradeoff}
\end{figure}

Figure \ref{fig:tradeoff} shows the trade-off between performance (represented through the balanced accuracy metric) and cost (as average inference time and total CPU time). Balanced accuracy is plotted against per-sample average inference time (log scale), while bubble size represents the total CPU time consumed by each method. 

\subsection{Impact of Migration}

Outside of the main RQs, we repeated the experiments using an API-based deployment of Open-Mistral-7B on remote infrastructure to assess the impact of local CPU-based deployment, since LLMs are typically deployed on GPU-accelerated environments. The results are summarized in Table~\ref{tab:mistral_api}.

In the table, we refer to the average processing time per sample as \emph{average latency}, as it includes both model inference and network overhead. Similarly, \emph{total time} refers to the elapsed wall-clock time from the start to the end of the prompting process. We also report the average number of tokens processed per request,  which is related to inference workload and is used as a proxy for deployment cost and efficiency in API-based deployments \cite{argerich2024measuring}.

As expected, the migration significantly reduced the end-to-end latency of the model. The API-based FS approach introduced only minimal overhead and slightly outperformed the local FS setup, though it remained below the performance of local ZS. Additionally, API-based ZS performed worse than both local approaches.

It is important to note that, unlike the local experiments, the API-based setting does not provide visibility into the actual computational resources consumed during inference (GPU utilization, power consumption, underlying hardware configuration). As a result, the true computational and energy costs associated with the deployment remain unknown.

\begin{table}
\centering
\caption{API-based deployment results for Open-Mistral-7B.}
\label{tab:mistral_api}
\begin{tblr}{
  row{1} = {c,font=\bfseries},
  row{2} = {font=\bfseries},
  column{2} = {c},
  column{3} = {c},
  column{5} = {c},
  column{6} = {c},
  cell{1}{1} = {c=3}{},
  cell{1}{4} = {c=3}{},
  cell{3}{4} = {r=3}{},
  cell{3}{5} = {r=3}{},
  cell{3}{6} = {r=3}{},
  cell{6}{4} = {r=3}{},
  cell{6}{5} = {r=3}{},
  cell{6}{6} = {r=3}{},
  cell{9}{4} = {r=3}{},
  cell{9}{5} = {r=3}{},
  cell{9}{6} = {r=3}{},
  vline{2} = {1}{},
  vline{4} = {2-11}{},
  hline{1,12} = {-}{0.08em},
  hline{2} = {-}{},
  hline{3} = {-}{0.05em},
}
 Detection &      &      & Inference       &       &       \\
Metric     & ZS   & FS   & Metric          & ZS    & FS    \\
TN         & 1    & 31   & Avg Latency (s) & 0.17  & 0.20  \\
FP         & 117  & 87   &                 &       &       \\
FN         & 8    & 53   &                 &       &       \\
TP         & 174  & 129  & Total Time (s)  & 50.82 & 59.35 \\
Acc        & 0.58 & 0.53 &                 &       &       \\
Bal Acc    & 0.48 & 0.49 &                 &       &       \\
Prec       & 0.60 & 0.69 & Avg Tokens      & 55    & 170   \\
Rec        & 0.96 & 0.71 &                 &       &       \\
F1         & 0.74 & 0.65 &                 &       &       
\end{tblr}
\end{table}

\section{Discussion}
\label{sec:disc}
The results show that traditional ML models significantly outperform LLM-based approaches for tabular 5G intrusion detection in terms of detection performance, inference time, and computational cost. Furthermore, few-shot prompting increases CPU time without providing consistent improvements in detection performance. On the contrary, the model exhibited a strong bias towards the malicious class, systematically flagging every sample as an anomaly. 

To verify that the weaker LLM results were not simply due to the reduced test set, we also evaluated XGBoost and TabNet on the same stratified subset of 300 samples used for the LLM experiments. Both models again achieved near-perfect detection performance, consistent with their results on the full test set. 

The substantially higher inference time of LLM-based approaches makes them impractical for real-time intrusion detection scenarios, where timely responses are essential. In this context, XGBoost provides the best overall trade-off, achieving near-perfect detection performance while maintaining minimal inference time and low computational demand.

Although ML models require an initial training phase, as shown in the experiments, this cost is relatively low and incurred offline, typically only once or infrequently. In contrast, LLMs rely on extensive pre-training on massive datasets using large-scale distributed clusters. For instance, training a model such as GPT-3 (175 billion parameters) would take a single NVIDIA A100 GPU approximately 32 years, while even a cluster of 1,024 GPUs requires over a month of training \cite{liu2025LLMs}. This highlights the substantial computational and infrastructural demands associated with modern LLMs.

While this cost was not considered in our study, the results show that LLMs introduce significant inference overhead, as each prediction requires processing a full prompt. These findings suggest that, for tabular 5G intrusion detection, tree ensemble methods remain a more efficient and reliable choice than current LLM-based approaches.

To assess whether the LLM's weak results are due to local deployment constraints, we conducted additional experiments using an API-based deployment of the same model on remote GPU infrastructure. While this setup reduced inference time to approximately 170-200 ms per sample, depending on the prompting strategy, it remained several orders of magnitude slower than traditional ML models. 

Mistral was selected to give LLMs a competitive chance, as it exhibits a balanced performance-to-efficiency ratio \cite{ferrag2025generative}. It is reasonable to assume that the results do not reflect a limitation of this specific LLM, and other general-purpose models would suffer from a similar trade-off. Even if the detection performance of other LLMs were to be superior, the computational cost would still be unfavorable.
 
Although this study is limited to a single dataset and a single LLM architecture, it provides an initial insight about the performance gap between LLMs and classic ML approaches for tabular data based intrusion detection. Additionally, LLM performance is highly sensitive to prompt design. Careful prompt engineering was necessary to constrain the model to produce consistent classification outputs and avoid undesired behaviors (e.g., generating verbose explanations or deviating from the task), which introduces an additional source of variability in the results.

It is also worth mentioning that this work focuses on supervised intrusion detection, where labeled malicious samples are available during training. While this setting enables effective detection of known attack patterns, it may be less suitable for identifying previously unseen threats, which are common in real-world deployments. In such scenarios, anomaly detection approaches can provide a complementary solution. 

\section{Conclusion and Future Directions}
\label{sec:conclusion}
This paper investigated the trade-offs between detection performance and computational cost in AI-based 5G intrusion detection. We compared two specialized models for tabular data, XGBoost and TabNet, with a general-purpose LLM, Open-Mistral-7B, evaluated under zero-shot and few-shot prompting.

The results show that traditional ML models achieve near-perfect detection performance while maintaining negligible inference time and substantially lower computational cost. In contrast, the LLM-based approach performs significantly worse and incurs several orders of magnitude higher inference time and CPU usage. Few-shot prompting further increases the computational cost without improving the overall trade-off. Additional experiments with API-based deployment indicate that improved hardware reduces inference time, but still remains slower than classic approaches.

Our findings confirm the initial hypothesis that LLMs are not the optimal solution for 5G intrusion detection, and suggest that tree ensemble models are a more practical and effective choice than both DNNs and LLMs in this context. 

This highlights the importance of selecting models based on task suitability rather than following general trends, and cautions against adopting increasingly complex AI systems when simpler approaches already provide superior performance and cost-efficiency.

It also remains unclear on which basis the LLM was discriminating between samples and why the few-shot prompting introduced such a strong bias towards the malicious class. This "black box" problem, which also affects deep learning models, is another open challenge in applying this technology to high-risk domains, such as cybersecurity.

It may be argued that LLMs do not require extensive training or specialized knowledge on the user side. However, pre-training LLMs remains one of the major challenges in terms of time and resource usage, as discussed above. In addition, due to the sensitivity to prompt engineering, domain knowledge is still necessary to some extent, as the model can make mistakes and hallucinate \cite{huang2025survey}.

Future work should extend this study in several directions. First, evaluating the proposed analysis in a fully unsupervised or semi-supervised setting, where malicious data is not available during training, would provide a more inclusive assessment of performance–cost trade-offs. 

Second, further research could explore LLMs specifically designed for tabular data, although current implementations remain limited to relatively simple tasks and are not yet suitable for intrusion detection\footnote{\url{https://github.com/SpursGoZmy/Awesome-Tabular-LLMs}}, and fine-tuned Small Language Models (SLMs), which offer reduced computational requirements and task-specific optimization capabilities \cite{wang2025comprehensive}.

Finally, investigating the impact of domain-specific fine-tuning and specialized knowledge bases could further improve the effectiveness of LLM-based approaches in cybersecurity applications.

\bibliographystyle{IEEEtran}
\bibliography{bib.bib}

\end{document}